\documentclass[aps,prc,twocolumn,groupedaddress,showpacs,10pt,dvipsnames]{revtex4-1}
\usepackage[utf8]{inputenc}
\usepackage{amsfonts}
\usepackage{amsmath}
\usepackage{amssymb}
\usepackage[pdftex]{graphicx}
\usepackage{xcolor}
\usepackage{tikz}
\usepackage{cancel}
\usetikzlibrary{arrows,patterns}

\begin{document}

\newcommand{\tj}[6]{ \begin{pmatrix}
  #1 & #2 & #3 \\
  #4 & #5 & #6
 \end{pmatrix}}

%\preprint{}

\title{Astrophysical reaction rate for $^9$Be formation within a three-body approach} 



%\thanks{}

\author{J. Casal}
\email{jcasal@us.es}
\author{M. Rodr\'{\i}guez-Gallardo}
\author{J. M. Arias}
\affiliation{Departamento de F\'{\i}sica At\'omica, Molecular y Nuclear,
  Facultad de F\'{\i}sica, Universidad de Sevilla, Apartado 1065, E-41080
  Sevilla, Spain} 
\author{I. J. Thompson}
\affiliation{Lawrence Livermore National Laboratory, L-414, Livermore, California 94551, USA}


\date{\today}

\begin{abstract}
The structure of the Borromean nucleus $^9$Be ($\alpha+\alpha+n$) is addressed within a three-body approach using the analytical transformed harmonic oscillator method. The three-body formalism provides an accurate 
description of the radiative capture reaction rate for the entire temperature range relevant in Astrophysics. At high temperatures, results match the calculations based on two-step sequential processes. At low temperatures,  where the particles have no access to intermediate two-body resonances, the three-body direct capture leads to reaction rates larger than the sequential processes.  These results support the reliability of the method for systems with several charged particles.
\end{abstract}

%\keywords{}

\maketitle

  %\section{}
  %\subsection{}
  %\subsubsection{}

  \section{Introduction}

The origin of elements in the Universe is an important topic in Nuclear Astrophysics~\cite{Hoyle}. The formation of heavy nuclei from light elements needs to overcome the instability gaps at 
mass numbers $A=5$ and $A=8$~\cite{Aprahamian05}. At the helium burning stage of stars, the triple-$\alpha$ reaction for the formation of $^{12}$C is the main nucleosynthesis process. 
However, in neutron rich environments, the reaction $\alpha(\alpha n,\gamma)^9\text{Be}$ followed by $^9\text{Be}(\alpha,n)^{12}\text{C}$ may dominate over, depending on the astrophysical 
conditions~\cite{Sumiyoshi02}. The relevance of this process has been linked to the nucleosynthesis by rapid neutron capture (or $r$ process) in type II 
supernovae~\cite{Efros98,Sumiyoshi02,Burda10,Arnold12}, so establishing an accurate rate for the formation of $^9$Be is essential for the $r$-process abundance predictions~\cite{Mengoni00,Sasaki05}.

The radiative three-body capture processes are essential in overcoming the $A=5,8$ gaps~\cite{Aprahamian05,Efros96}, but traditionally they have been described as two-step sequential 
reactions~\cite{Fowler67,Hoyle,Angulo99,Sumiyoshi02,Grigorenko06,Alvarez08}. When at least one of the two-body subsystems shows a low-lying narrow resonance, the sequential picture provides 
a rather accurate description of these reactions for high-temperature environments, where the intermediate states can be populated. However, at low temperatures the particles may have no 
energy to populate intermediate resonances, and therefore the direct three-body capture plays an important role~\cite{Garrido11,RdDiego10,Nguyen1}. Moreover, the intermediate 
configurations may not be present or show a too quick decay. So, a complete three-body formulation is needed to describe properly the reaction rates of such nuclei in the entire temperature range. 

The complete computation of three-body reactions in the whole energy range requires a narrow grid of continuum states right above the breakup threshold~\cite{Garrido11}, which is a 
difficult task.
The asymptotic behavior of continuum states for systems with several charged particles is not known in general, and very 
involved procedures are needed to deal with this problem~\cite{Nguyen1,Nguyen2,Ishikawa13}. In a recent work~\cite{JCasal13,JCasalAIP12} we presented a pseudostate (PS) method based on an analytical local scale transformation (LST) of 
the harmonic oscillator (HO) basis, the transformed harmonic oscillator (THO) method. We generalized the analytical THO method for three-body systems and successfully applied to the Borromean nucleus $^6$He ($\alpha+n+n$) 
system. PS methods consist in diagonalizing the Hamiltonian in a complete set of square-integrable functions, a procedure which does not require going through the continuum 
wave functions, and the previous knowledge of the asymptotic behavior is not needed. 
Furthermore, in the analytical THO method, the parameters
of the transformation govern the radial extension of the THO
basis. This provides the advantage of  allowing the construction of an optimal basis for each observable of interest~\cite{JALay10,AMoro09,JCasal13}.
The analytical THO basis can describe very accurately the strength functions  in the low-energy range,  providing a good description of the radiative capture reactions. 

In the present work we apply the analytical THO method to the Borromean nucleus $^9$Be, whose astrophysical relevance has been pointed out. The purpose of this paper is to 
show the reliability of the method when applied to systems with more than one charged particle, and to confirm the importance of the direct three-body capture at low temperature.  The full 
three-body formalism allows the treatment of the direct and sequential, resonant and non-resonant processes in the same footage. Thus these processes do not need to be treated separately when estimating 
the total contribution to the astrophysical reaction rate~\cite{Alvarez07,Nguyen1}.

The paper is structured as follows. In Sec.~\ref{sec:THO} the three-body formalism is presented. The analytical THO method and the expression for  the radiative capture reaction rate are shown.  The electromagnetic transition probabilities are derived for a system with two identical charged particles.  
In Sec.~\ref{sec:application} the full formalism is applied to the case of $^9$Be, and the rate of the radiative capture reaction $\alpha+\alpha+n\rightarrow$ $^9\text{Be} +\gamma $ is obtained.
Finally, in Sec. \ref{sec:conclusions}, the main conclusions of this work are summarized.

\section{Three-body formalism}
\label{sec:THO}

The three-body formalism used in this work is described in detail in Ref.~\cite{JCasal13}, where it is applied to a system with a single charged particle. In this section,  we summarize the main features of the formalism and we derive the electromagnetic transition probabilities $B(\mathcal{O}\lambda)$ 
for the case of a system with two identical charged particles, such as $^{9}$Be $(\alpha + \alpha + n)$.

%\tikzset{>=latex}
\begin{figure}
\centering

% \begin{tikzpicture}[scale=0.35]
%  \draw [fill=gray!10] (0,0) circle (20pt);
%  \draw [fill=gray!40] (1,4) circle (15pt);
%  \draw [fill=gray!90] (4,1) circle (10pt);
%  \draw [<-] (0,0) -- (4,1);
%  \draw [->] (1.45,0.44) -- (1,4);
%  \node [above right] at (1,4.3) {\small $1$};
%  \node [below right] at (4.3,1) {\small $2$};
%  \node [below left] at (-0.3,0) {\small $3$};
%  \node [below right] at (1.88235,0.47059) {\small $\vec{x}_1$};
%  \node [above right] at (1.5,2) {\small $\vec{y}_1$};
% \end{tikzpicture}\hspace{0.5cm}
% \begin{tikzpicture}[scale=0.35]
%  \draw [fill=gray!10] (0,0) circle (20pt);
%  \draw [fill=gray!40] (1,4) circle (15pt);
%  \draw [fill=gray!90] (4,1) circle (10pt);
%  \draw [->] (0,0) -- (1,4);
%  \draw [->] (0.5,1.6) -- (4,1);
%  \node [above right] at (1,4.3) {\small $1$};
%  \node [below right] at (4.3,1) {\small $2$};
%  \node [below left] at (-0.3,0) {\small $3$};
%  \node [left] at (0.53846,2.15385) {\small $\vec{x}_2$};
%  \node [above] at (2.5,1.7) {\small $\vec{y}_2$};
% \end{tikzpicture}\hspace{0.5cm}
%  \begin{tikzpicture}[scale=0.35]
%  \draw [fill=gray!10] (0,0) circle (20pt);
%  \draw [fill=gray!40] (1,4) circle (15pt);
%  \draw [fill=gray!90] (4,1) circle (10pt);
%  \draw [->] (1,4) -- (4,1);
%  \draw [->] (2.2,2.8) -- (0,0);
%  \node [above right] at (1,4.3) {\small $1$};
%  \node [below right] at (4.3,1) {\small $2$};
%  \node [below left] at (-0.3,0) {\small $3$};
%  \node [above right] at (2.5,2.5) {\small $\vec{x}_3$};
%  \node [above left] at (1.2,1.2) {\small $\vec{y}_3$};
% \end{tikzpicture}

 \includegraphics[width=\linewidth]{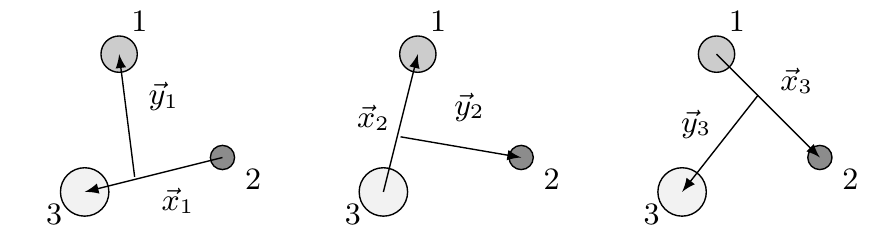}
 \caption{The three sets of scaled Jacobi coordinates.}
 \label{fig:sets}
\end{figure}

In order to describe the three-body system we use hyperspherical coordinates $\{\rho,\alpha_k,\widehat{x}_k,\widehat{y}_k\}$, which are obtained from the Jacobi coordinates 
$\{\boldsymbol{x}_k,\boldsymbol{y}_k\}$. Note that there are three possible Jacobi systems, each one denoted by the label $k=1,2,3$.  The variable $\boldsymbol{x}_k$ is proportional to the
relative coordinate between two of the particles and $\boldsymbol{y}_k$
is proportional to  
the coordinate from the center of mass of these two particles to the
third one, both with a scaling factor depending on their
masses~\cite{MRoGa05}. 
We are using the odd man out notation in which, for example, the Jacobi-$1$ system corresponds to the Jacobi system in which the particles (2,3) are related by the coordinate $\boldsymbol{x}_1$ (see Fig.~\ref{fig:sets}).

The hyper-radius $\rho$ and the hyperangle $\alpha_k$ are related to 
the Jacobi coordinates as
\begin{align}
\rho = & \sqrt{x_k^2 + y_k^2}, \\
\alpha_k = & \tan\left(\frac{x_k}{y_k}\right).
\label{eq:jachyp}
\end{align}
While the hyperangle depends on the Jacobi-$k$ system, the hyper-radius does not. 

\subsection{Analytical THO}

The THO method consists in diagonalizing the Hamiltonian of the system in a discrete basis of $\mathcal{L}^2$ functions, the THO functions in one of the Jacobi system (for simplicity, if $k$ is fixed we do not specify it)   
\begin{equation}
 \psi^{\text{THO}}_{i\beta j\mu} (\rho,\Omega) = R^{\text{THO}}_{i\beta}(\rho) \mathcal{Y}_{\beta j\mu}(\Omega),
\label{eq:basis}
\end{equation}
where $\Omega\equiv\{\alpha,\widehat{x},\widehat{y}\}$ is introduced for the angular dependence and $\beta\equiv\{K,l_x,l_y,l,S_x,j_{ab}\}$ is a set of quantum numbers called channel. In this set, $K$ is the hypermomentum, $l_x$ and  
$l_y$ are the orbital angular momenta associated with the Jacobi
coordinates $\boldsymbol{x}$ and $\boldsymbol{y}$, respectively, $l$  is the
total orbital angular momentum
($\boldsymbol{l}=\boldsymbol{l_x}+\boldsymbol{l_y}$), $S_x$ is the
spin of the particles related by the coordinate $\boldsymbol{x}$, and
$j_{ab}$ results from the coupling
$\boldsymbol{j_{ab}}=\boldsymbol{l}+\boldsymbol{S_x}$. If we denote by
$I$  the spin of the third particle, that we assume to be fixed, the
total angular momentum $j$ is $\boldsymbol{j}=\boldsymbol{j_{ab}} +
\boldsymbol{I}$.
The functions $\mathcal{Y}_{\beta j\mu}(\Omega)$ 
are states of good total angular momentum, expanded in hyperspherical harmonics (HH)~\cite{Zhukov93,MRoGaTh} as shown in the Appendix (see Eq.~(\ref{eq:HHexpand})).

The THO hyper-radial functions $R^{\text{THO}}_{i\beta}(\rho)$ are based on a LST, $s(\rho)$, of the HO functions
\begin{equation}
  R_{i\beta}^{\text{THO}}(\rho)=\sqrt{\frac{ds}{d\rho}}R_{iK}^{\text{HO}}[s(\rho)],
\label{eq:R}
\end{equation}
where $i$ denotes the hyper-radial excitation. In this paper, as in Refs.~\cite{JCasal13,JALay10,AMoro09}, we adopt the analytical form of Karataglidis 
\textsl{et al.}~\cite{Karataglidis},
\begin{equation}
s(\rho) = \frac{1}{\sqrt{2}b}\left[\frac{1}{\left(\frac{1}{\rho}\right)^{\xi} +
\left(\frac{1}{\gamma\sqrt{\rho}}\right)^\xi}\right]^{\frac{1}{\xi}},
\label{eq:LST}
\end{equation}
depending on the parameters $\xi$, $\gamma$, and the oscillator length $b$. The HO hyper-radial variable $s$ is dimensionless according to the transformation defined above 
[Eq.~(\ref{eq:LST})]. In this way, we take the oscillator length $b$ as another parameter of the transformation. We have fixed for all calculations $\xi=4$ as in Ref.~\cite{JCasal13}, since it was found previously a very weak dependence of the results on this parameter. 
Note that the THO hyper-radial wave functions depend, in general, on all the quantum numbers included in a channel $\beta$, however the HO 
hyper-radial wave functions only depend on the hypermomentum $K$.

The states of the system are then given by
diagonalization of the three-body Hamiltonian in a finite basis up to a maximum hypermomentum $K_{\rm max}$, which determines the number of channels, and $i_{\rm max}$
hyper-radial excitations in each channel,
\begin{equation}
 \Psi_{nj\mu}(\rho,\Omega)=\sum_{\beta}\sum_{i=0}^{i_{\rm max}} C_n^{i\beta j} R_{i\beta}^\text{THO}(\rho)\mathcal{Y}_{\beta j\mu}(\Omega),
 \label{eq:wf}
\end{equation}
where $C_n^{i\beta j}$ are the diagonalization coefficients, and the label $n$ enumerates the eigenstates.

The function $s(\rho)$ behaves asymptotically as $\frac{\gamma}{b}\sqrt{\frac{\rho}{2}}$ and hence the THO hyper-radial wave functions obtained behave at large distances as 
$\exp{(-\gamma^2\rho/2b^2)}$. Therefore, the ratio $\gamma/b$ governs the asymptotic behavior of the THO functions: as $\gamma/b$ increases, the hyper-radial extension of the basis 
decreases and some of the eigenvalues obtained by diagonalizing the Hamiltonian explore higher energies~\cite{JALay10}. That is, $\gamma/b$ determines the density of PSs as a function of 
the energy. This gives the freedom to choose an appropriate basis depending on the observable of interest.

  \subsection{Radiative capture reaction rate}\label{sec:rate}

We consider the radiative capture reaction of three particles, ($abc$), into a bound nucleus $A$ of binding energy $|\varepsilon_B|$, \textsl{i.e.} \mbox{$a+b+c \rightarrow$ $A + \gamma$}. 
The energy-averaged reaction rate for such process, $\langle R_{abc}(\varepsilon)\rangle$, is given as a function of the temperature  $T$ by the expression~\cite{Garrido11,JCasal13}
\begin{align}
 \nonumber\langle R_{abc}(\varepsilon) \rangle(T) = & ~\nu!\frac{\hbar^3}{c^2}\frac{8\pi}{(a_x a_y)^{3/2}}\frac{g_A}{g_a g_b g_c} \frac{1}{(k_B T)^3}  \\
 \times & \int_0^\infty (\varepsilon+|\varepsilon_B|)^2 \sigma_\gamma^{(\mathcal{O}\lambda)}(\varepsilon+|\varepsilon_B|) e^{\frac{-\varepsilon}{k_B T}} d\varepsilon.
 \label{eq:aRE}
\end{align}
where $\varepsilon=\varepsilon_\gamma+\varepsilon_B$ is the initial three-body kinetic energy, $\varepsilon_\gamma$ is the energy of the photon emitted, $\varepsilon_B$ is the ground-state 
energy, $g_i$ are the spin degeneracy of the particles, $\nu$ is the number of identical particles in the three-body system, $a_x$ and $a_y$ are the reduced masses of the subsystems 
related to the Jacobi coordinates $\{\boldsymbol{x},\boldsymbol{y}\}$, and $\sigma_\gamma(\varepsilon_\gamma)$ is the photodissociation cross section of the system $A$. This function 
can be expanded into electric and magnetic multipoles~\cite{RdDiego10,Forseen03}
\begin{equation}
\sigma_\gamma^{(\mathcal{O}\lambda)}(\varepsilon_\gamma)=\frac{(2\pi)^3 (\lambda+1)}
{\lambda[(2\lambda+1)!!]^2}\left(\frac{\varepsilon_\gamma}{\hbar
c}\right)^{2\lambda-1}\frac{dB(\mathcal{O}\lambda)}{d\varepsilon},
\label{eq:xsection}
\end{equation}
 which are related to the transition probability distributions
$dB(\mathcal{O}\lambda)/d\varepsilon$, for $\mathcal{O}=E, M$. 

The integral in Eq.~(\ref{eq:aRE}) is very sensitive to the $dB(\mathcal{O}\lambda)/d\varepsilon$ behavior at low energy and, for that reason, a detailed description of the transition 
probability distribution in that region is needed to avoid numerical errors.  Accordingly to the traditional literature~\cite{Weiss}, in the absence of low energy resonances the first multipole 
contribution is the dominant one and the electric contribution dominates over the magnetic one at the same order.

    \subsection{Electromagnetic transition probability $\boldsymbol{B(\mathcal{O}\lambda)}$}\label{ss:bel}

As in Refs.~\cite{JCasal13,MRoGa05}, we follow the notation of Brink and Satchler~\cite{BrinkSatchler}. The reduced transition probability between states of the system is defined as 
\begin{eqnarray}
 \nonumber B(\mathcal{O}\lambda)_{nj,n'j'} & \equiv & B(\mathcal{O}\lambda;nj\rightarrow n'j') \\
 & =& |\langle nj\|\widehat{\mathcal{O}}_\lambda\|n'j'\rangle|^2\left(\frac{2\lambda+1}{4\pi}\right),
\label{eq:BE}
\end{eqnarray}
where $\widehat{\mathcal{O}}_{\lambda M_\lambda}$ is the electric or magnetic multipole operator of order $\lambda$, and the $|nj\mu\rangle$ denotes the wave function given by Eq.~(\ref{eq:wf}).

We consider first electric transitions, involving the matrix elements of the electric multipole operator $\widehat{Q}_{\lambda M_\lambda}$. This operator, for a general system with three particles, takes the form in the Jacobi-$k$ set
\begin{equation}
\widehat{Q}_{\lambda M_{\lambda}}(\boldsymbol{x}_k,\boldsymbol{y}_k)=\left(\frac{4\pi}{2\lambda+1}\right)^{1/2}\sum_{q=1}^3 Z_q~e~r_q^\lambda Y_{\lambda M_{\lambda}}(\widehat{r}_q),
\label{eq:Qop}
\end{equation}
where $Z_q$ is the atomic number of the particle $q$, $e$ is the electron charge, and $\boldsymbol{r}_q$ is the position of particle $q$ with respect to the center of mass of the system, which in the 
Jacobi-$q$ system is given by~\cite{Nielsen01} 
\begin{equation}
 \boldsymbol{r}_q = \sqrt{\frac{m}{m_q}\frac{\left(M_T-m_q\right)}{M_T}}\boldsymbol{y}_q.
 \label{eq:relpos}
\end{equation}
Here $m$ is a normalization mass, taken as the atomic mass unit, and $M_T$ is the total mass of the system. We describe the system in a preferred Jacobi set, $k$, however the expression for the electric multipole operator given by Eq.~(\ref{eq:BE}) can be easily 
expressed, in general, using different Jacobi systems. The relation between  harmonic polynomials in different Jacobi sets is given by the expression~\cite{RdDiego08}
\begin{align}
 y_q^\lambda Y_{\lambda M_\lambda}\left(\widehat{y}_q\right)  = & \sum_{l=0}^{\lambda}\left(-1\right)^\lambda x_k^{\lambda-l}\left(\sin\varphi_{qk}\right)^{\lambda-l} y_k^l \left(\cos\varphi_{qk}\right)^l \nonumber\\
& \times \sqrt{\frac{4\pi\left( 2\lambda+1\right)!}{\left( 2l+1\right)!\left( 2\lambda-2l+1\right)!}} \nonumber \\
& \times \left[Y_{\lambda-l}\left(\widehat{x}_k\right) \otimes Y_{l}\left( \widehat{y}_k\right) \right]^{\lambda M_\lambda},
 \label{eq:harmonicp}
 \end{align}
with
 \begin{equation}
 \tan\varphi_{qk} = \left( -1\right)^P\sqrt{\frac{m_p M_T}{m_q m_k}}~,
 \label{eq:phi}
 \end{equation}
depending on the mass of the particles and the parity $(-1)^P$ of the permutation $P$ of $\{k,p,q\}$. The identity transformation is given by $\varphi_{kk}=\pi$.
Using Eq.~(\ref{eq:harmonicp}) we can rewrite the harmonic polynomial for each particle $q$, as a function of the Jacobi coordinates in the preferred Jacobi system $k$. 
This is equivalent to rotating the functions to the Jacobi system $q$ where the position of each particle is given by a vector proportional to $\boldsymbol{y}_q$. 

\begin{figure}
\centering

% \begin{tikzpicture}[scale=0.35]
%  \draw [fill=Blue!60] (0,0) circle (10pt);
%  \draw [fill=cyan!60] (1,4) circle (25pt);
%  \draw [fill=cyan!60] (4,1) circle (25pt);
%  \draw [->] (1,4) -- (4,1);
%  \draw [->] (2.5,2.5) -- (0,0);
%  \node [above right] at (1.2,4.5) {\small $\alpha$};
%  \node [above left] at (1.2,4.8) {\small 1};
%  \node [above right] at (4.2,1.55) {\small $\alpha$};
%  \node [below right] at (4.8,1.45) {\small 2};
%  \node [above left] at (0,0.3) {\small $n$};
%  \node [below left] at (-0.4,0.1) {\small 3};
%  \node [above right] at (2.5,2.5) {\small $\vec{x}$};
%  \node [above left] at (1.2,1.2) {\small $\vec{y}$};
% \end{tikzpicture} 

 \includegraphics[width=0.35\linewidth]{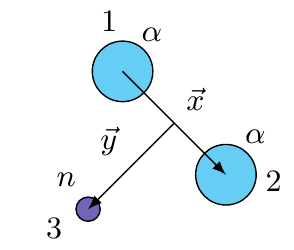}
 \caption{(Color online) $^9$Be in Jacobi-$T$ system.}
 \label{fig:Tsys}
\end{figure}

If we consider a system with two identical charged particles, such as $^9$Be, we describe the problem using the Jacobi $T$-system shown in Fig.~\ref{fig:Tsys}. In the T-system the two $\alpha$ particles are related by the $\boldsymbol{x}$ coordinate. For simplicity, the subindexes corresponding to the chosen Jacobi set are normally omitted. So in this case, $\boldsymbol{x}$=$\boldsymbol{x}_3$ and $\boldsymbol{y}$=$\boldsymbol{y}_3$.

From Eqs.~(\ref{eq:relpos}), (\ref{eq:harmonicp}), 
and (\ref{eq:phi}), the expression (\ref{eq:Qop}) can be reformulated for dipolar transitions $\left(\lambda=1\right)$ as
\begin{equation}
 \widehat{Q}_{1M_1} = -\left(\frac{4\pi}{3}\right)^{1/2} 2\left(\cos\varphi_{23}\right) Z_2e\frac{\sqrt{ma_{y2}}}{m_2}yY_{1M_1}\left(\widehat{y}\right).
 \label{eq:Qop1_9be}
\end{equation}
Here $a_{y2}=a_{y1}$ is the Jacobi mass factor related to the coordinate $\boldsymbol{y}_2$, 
\begin{equation}
 a_{y2}=\frac{m_2\left(m_3+m_1\right)}{M_T},
\end{equation}
and $m_2=m_1$, in this case, is the $\alpha$ particle mass. The $\boldsymbol{x}$ component is absent in Eq.~(\ref{eq:Qop1_9be}) because the two charged particles are identical, what simplifies the problem. This expression is analogous to Eq.~(18) in Ref.~\cite{JCasal13} but including a factor $2\cos\left(\varphi_{23}\right)$ which, for $^9$Be, equals $2/\sqrt{10}$. 

To test the completeness of the basis, we can also calculate the sum rule for electric dipolar transitions from the ground 
state (g.s.) to the states $(n,j)$. Using the equations (\ref{eq:BE}) and (\ref{eq:Qop1_9be}) we obtain
\begin{equation}
S_T(E1)=\frac{3}{4\pi}\frac{Z^2e^2m a_{y2}}{m_2^{2}} \left(2\cos\varphi_{23}\right)^2\langle\text{g.s.}| y^{2}|\text{g.s.}\rangle. 
 \label{eq:sumrule}
\end{equation}

If the system shows low-energy resonances coupled to the ground state by magnetic transitions at the same order than electric transitions, magnetic contributions may play a significant role.
We consider then magnetic transitions, involving the matrix elements of the magnetic operator $\widehat{M}_{\lambda M_\lambda}$. This operator can be expressed as a sum of two terms, the orbital and spin terms \cite{BM}. Following the notation of Brink and Satchler
\begin{align}
 \widehat{M}^{\text{orb}}_{\lambda M_\lambda}(\vec{r})= & \frac{e\hbar}{2mc}\sqrt{4\pi\lambda}\sum_q r_q^{\lambda-1}\frac{2g_l^{(q)}}{\lambda+1}\left[Y_{\lambda-1}l\right]_{(\lambda-1,1)\lambda,M_\lambda}^{(q)},\\
 \widehat{M}^{\text{spin}}_{\lambda M_\lambda}(\vec{r})= & \frac{e\hbar}{2mc}\sqrt{4\pi\lambda}\sum_q r_q^{\lambda-1} g_s^{(q)}\left[Y_{\lambda-1}s\right]_{(\lambda-1,1)\lambda,M_\lambda}^{(q)}.
 \label{eq:Mop}
\end{align}
Here $g_l$ and $g_s$ are the orbital and spin $g$-factors, and $\left[Y_{\lambda-1}j\right]_{(\lambda-1,1)\lambda M_\lambda}$ is a tensorial product of order one,
\begin{equation}
 \begin{split}
 \left[Y_{\lambda-1}j\right]_{(\lambda-1,1)\lambda,M_\lambda}\equiv & \sum_{\eta\nu} Y_{(\lambda-1)\eta}\sqrt{2{j}_\nu+1}\\
                                                                    & \times\langle(\lambda-1)\eta 1\nu|\lambda M_\lambda\rangle.
 \label{eq:tensprod}
 \end{split}
\end{equation}
For dipolar transitions, the total magnetic operator is given then by 
\begin{equation}
 \begin{split}
 \widehat{M}_{1M_1}&=\widehat{M}^{\text{orb}}_{1M_1}+\widehat{M}^{\text{spin}}_{1M_1}\\
                   &=\frac{e\hbar}{2mc}\sum_q\left[g_l^{(q)}\boldsymbol{l}_q + g_s^{(q)}\boldsymbol{s}_q\right]_{M_1}
 \label{eq:Mop1}
 \end{split}
\end{equation}
These terms need to be evaluated for each particle. We express again the position of particle $q$ in the Jacobi-$q$ system by Eq.~(\ref{eq:relpos}), and we rotate the wave functions $|nj\rangle$ to that system using the transformations 
between different Jacobi sets (see, for instance, Ref.~\cite{IJThompson04}). The matrix element formula is given in Appendix~\ref{appendix:1}

Transition probabilities  given by Eq.~(\ref{eq:BE}) are a set of discrete values. In order
to obtain a continuous energy distribution, the best option is to do the overlap with the continuum wave functions~\cite{MRoGa11}, which are not known in general. 
In this work, as in Ref.~\cite{MRoGa05}, we  consider that a PS with energy $\varepsilon_n$ is the superposition of continuum states in the vicinity. There are several ways to assign an energy distribution to a PS. Here, as in Ref.~{\cite{JCasal13}} we 
assign a Poisson distribution for each discrete value of $B(\mathcal{O}\lambda)(\varepsilon_n)$, with the form 
\begin{equation}
  D(\varepsilon,\varepsilon_n,w)=\frac{(w+1)^{(w+1)}}{\varepsilon_n^{w+1}\Gamma(w+1)} \varepsilon^w\exp{\left(-\frac{w+1}{\varepsilon_n}\varepsilon\right)},
\label{eq:poisson}
\end{equation}
which is properly normalized. Poisson distributions tend smoothly to zero at the origin, which is the physical behavior we expect for the energy distributions of the pseudostates. The parameter $w$ controls the width of the 
distributions; as $w$ decreases, the width of the distributions increases.
The prescription to fix an appropriate $w$ parameter will be the same introduced in the former Ref.~\cite{JCasal13}. It consists in choosing the value of $w$ that ensures a smooth $B(E1)$ distribution without spreading
it unphysically. We present more details and a practical example in Appendix~\ref{appendix:2}.

  \section{Application to $^9$Be}\label{sec:application}
  The $^9$Be nucleus can be described as a three-body system, comprising two $\alpha$ particles and one neutron. It shows a Borromean structure, since none of the binary subsystems $^5$He nor 
$^8$Be are bound. $^9$Be is a loosely bound system with a 3/2$^-$ ground state located at 1.5736 MeV below the $\alpha + \alpha + n$ threshold~\cite{Tilley04}. The presence of a very narrow 
two-body $^8$Be resonance at 0.092 MeV above the three-body threshold suggests a sequential description of the formation process~\cite{Sumiyoshi02}. Due to the small lifetime of the $^5$He 
system ($\sim 10^{-21}$ s) compared to $^8$Be ($\sim 10^{-16}$ s), the sequential synthesis is considered to proceed mainly through $^8$Be~\cite{Utsunomiya01}. Nevertheless, the sequential 
picture may underestimate the reaction rate at low temperature by several orders of magnitude~\cite{Garrido11}.

  This nucleus presents a genuine three-body 1/2$^+$ resonant state around 0.11 MeV with a relatively large width~\cite{Garrido10}. Therefore, the photodissociation cross section of $^9$Be 
shows a relatively broad peak at the energy of the resonance, very close to the three-body and two-body thresholds. This resonance is the main contribution to the 
$\alpha(\alpha n,\gamma)^9\text{Be}$ reaction rate, especially at the lowest temperatures where other $j^\pi$ contributions are negligible~\cite{Arnold12,Garrido11}. The experimental cross section shows also a rather narrow peak around 0.85 MeV associated to the 
5/2$^-$ resonance. In this work, we have included in the calculation the 1/2$^+$, 3/2$^+$, 5/2$^+$ states, all connected to the ground state by electric dipolar (E1) transitions. Magnetic dipolar (M1) transitions to the 
1/2$^-$, 3/2$^-$, 5/2$^-$ states are also known to have an influence on the reaction rate~\cite{Arnold12,Utsunomiya01,Sumiyoshi02,Burda10}. Although they are not expected to change the 
low-temperature tail~\cite{Arnold12}, we have also calculated magnetic contributions. Our model treats the resonant and non-resonant parts of the spectrum in the same footage, 
both contributing to the strength function and the reaction rate.

 \subsection{Hamiltonian}
 Our three-body model includes the $\alpha$--$n$ potential from Ref.~\cite{IJThompson00}, which has been shown to provide reasonable results for $^6$He~\cite{MRoGa05,JCasal13}. In order to account for the Pauli principle needed to block occupied $\alpha$ states 
 to the neutron, a repulsive s-wave component is introduced in the $\alpha$--$n$ interaction, with the requirement that the  experimental phase shifts are correctly reproduced. For the $\alpha$--$\alpha$ nuclear interaction we include the Ali-Bodmer 
 potential~\cite{AliBodmer} version ``a'' with a different repulsive term for s- and d-waves, 
 \begin{equation}
  V_{\alpha\alpha}(r) = \left(125 \widehat{P}_{l=0} + 20 \widehat{P}_{l=2}\right)e^{-\left({r}/{1.53}\right)^2} - 30~ e^{-\left({r}/{2.85}\right)^2}.
 \end{equation}
 In this expression, the repulsive terms block the $\alpha$--$\alpha$ bound states, and their strengths need to be different in order to reproduce the experimental phase shifts. This potential together with a hard-sphere Coulomb interaction with a Coulomb 
 radius of $r_{\text{Coul}}=2.94$ fm,
 \begin{equation}
  V_{\alpha\alpha}^{Coul}(r) = Z^2e^2 \times 
  \begin{cases}
  \left(\frac{3}{2}-\frac{r^2}{2r_{Coul}^2}\right)\frac{1}{r_{Coul}} & r \leq r_{Coul} \\
  \frac{1}{r} & r > r_{Coul},
  \end{cases}
  \label{eq:coulombint}
 \end{equation}
 reproduces the exact position of the two-body s-wave $^8$Be resonance. The modification of the Ali-Bodmer potential introduced by Fedorov \textit{et al.}~\cite{Fedorov93} should not be used in combination with the Coulomb interaction given by Eq.~(\ref{eq:coulombint}), 
 since they do not reproduce the position of the two-body resonance, and this is crucial to obtain the right behavior in the low-lying $^{9}$Be continuum.
 
 These binary interactions are adjusted to reproduce the phenomenology of the two-body systems. Since three-body models are an approximation to the full many-body system, including only two-body interactions may lead to deviations from the experimental 
 three-body energies~\cite{IJThompson04,RdDiego10,MRoGa05}. Therefore, it is usual to include a structureless hyperradial three-body force, which can be fixed to adjust the energy of the system 
 without distorting its structure. We use the following expression, as in Refs.~\cite{MRoGa05,JCasal13},
 \begin{equation}
  V_{3b}(\rho) = \frac{v_{3b}}{1+\left(\frac{\rho}{r_{3b}}\right)^{a_{3b}}}~.
 \label{eq:3bf}
 \end{equation}
 There are different choices in the literature, and we have checked that the specific form of this interaction plays a negligible role on the final results. The parameters for the three-body force are chosen to adjust the energy of the experimentally known states 
 of the system, in this case, the ground  state of $^9$Be and the 1/2$^+$, 3/2$^+$, 5/2$^+$, 5/2$^-$ and 1/2$^-$ resonances. The value of these parameters are different for each $j^\pi$ state, and they are given in Table I. 
 % i.g. gaussian~\cite{Nguyen2} or potential~\cite{IJThompson04}%
 % We also include a simple hyperradial three-body force to account for the fact that the $\alpha$ particles are not fundamental, as in Refs.~\cite{JCasal13,Nguyen1},

\begin{table}
 \begin{tabular}{cccc} 
 \toprule
 $j^\pi$ & $v_{3b}$ (MeV) & $r_{3b}$ (fm) & $a_{3b}$ \\
 \colrule
 3/2$^-$ & $+$1.11 & 6.1 & 5\\ %1.1112
 1/2$^+$ & $-$2.45 & 6.1 & 5\\
 3/2$^+$ & $-$1.60 & 6.1 & 5\\
 5/2$^+$ & $-$0.18 & 6.1 & 5\\
 5/2$^-$ & $+$1.65 & 6.1 & 5\\
 1/2$^-$ & $+$0.20 & 6.1 & 5\\
 \botrule
 \end{tabular}
\caption{Three-body force (Eq. (\ref{eq:3bf})) parameters for different $j^\pi$ states. See text for details on $K_{max}$ and $i_{max}$ values for each $j^\pi$.}
\end{table}
%Calculations were performed with $K_{\rm max}=30$ and $i_{\rm max}=30$.

We diagonalize the Hamiltonian in a finite THO basis with maximum value of the hypermomentum $K_{\rm max}$ and a maximum number of hyperradial excitations in each channel $i_{\rm max}$. We calculate separately the kinetic energy matrix elements and the potential 
matrix elements. The hyperangular integration of the potential matrix elements are performed, as in Refs.~\cite{JCasal13,MRoGa05}, by using a set of subroutines of the code 
\textsc{face}~\cite{IJThompson04}.

 \subsection{3/2$^-$ ground state}

The 3/2$^-$ states are described with an analytical THO basis defined by parameters $b=0.7$ fm, and $\gamma=1.4$ fm$^{1/2}$, trying
to minimize the size of the basis needed to reach convergence
of the ground state.
The three-body force parameters are taken as $v_{3b}=1.11$ MeV, 
$r_{3b} = 6.1$ fm and $a_{3b}= 5$, chosen to adjust the ground-state energy and the matter radius of $^{9}$Be.

\begin{figure}
\includegraphics[width=\linewidth]{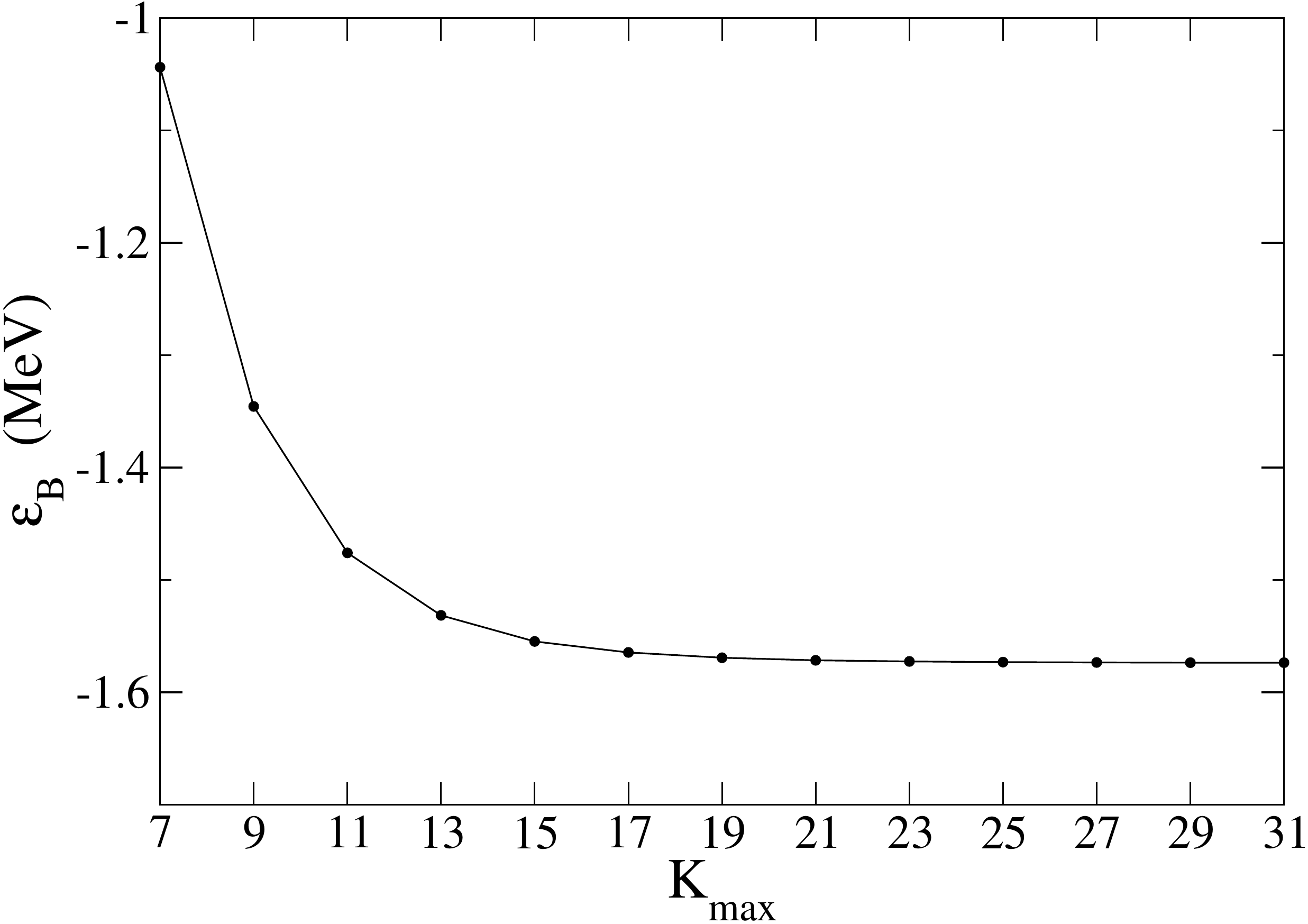}
\caption{Convergence of the ground-state energy of $^9$Be with respect to the maximum hypermomentum $K_{\rm max}$.}
\label{fig:ebkmax}
\end{figure}

\begin{figure}
\includegraphics[width=\linewidth]{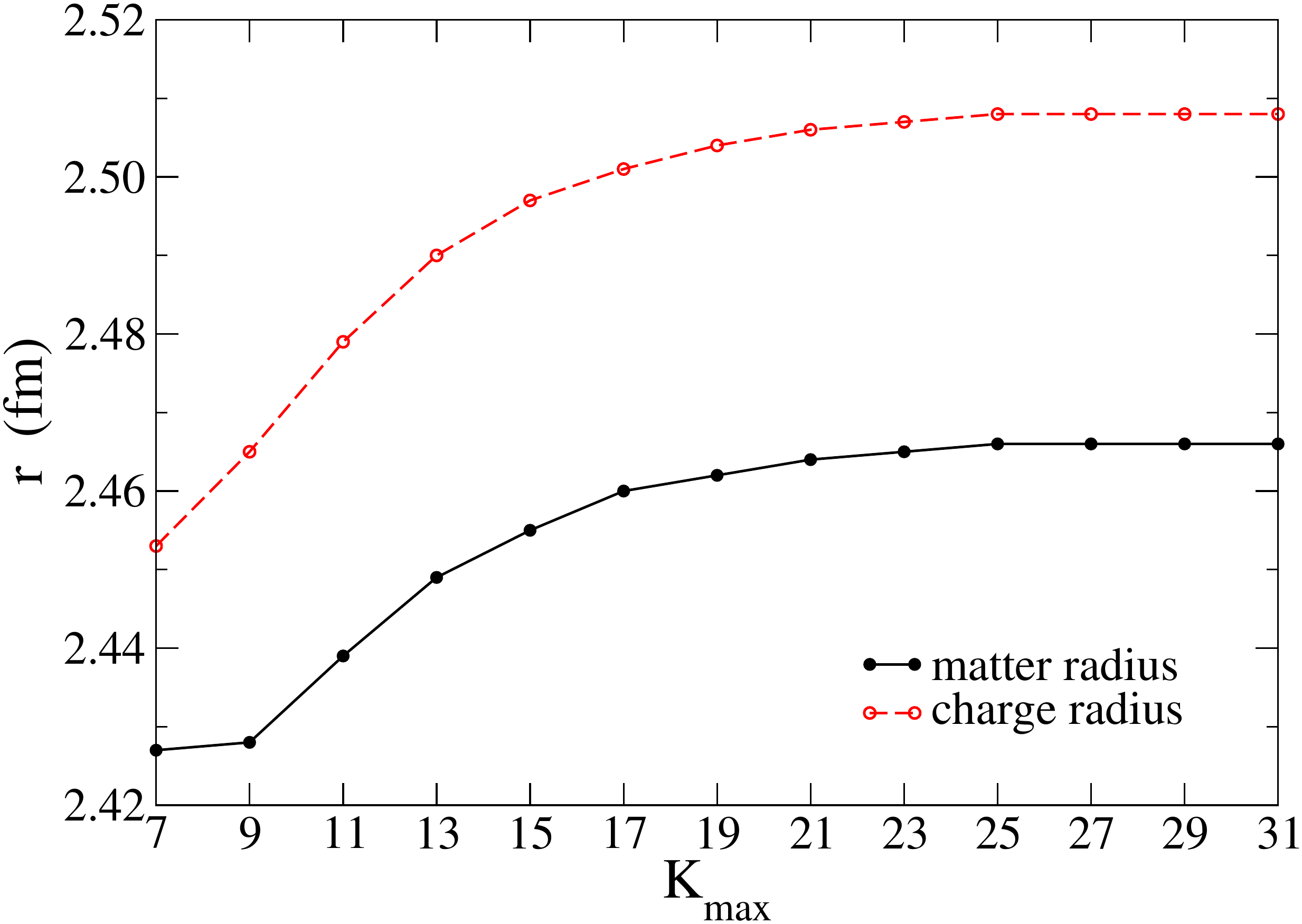}
\caption{(Color online) Convergence of the matter radius (solid line) and the charge radius (dashed line) of $^9$Be with respect to the maximum hypermomentum $K_{\rm max}$.}
\label{fig:rmatkmax}
\end{figure}

In Figs.~\ref{fig:ebkmax} and \ref{fig:rmatkmax} we show the convergence of the ground-state energy and the matter and charge radii with respect to the maximum hypermomentum $K_{\rm max}$ with $i_{\rm max}$ fixed to 20.  $K_{\rm max}$ determines 
the number of channels included in the wave function expansion. From Fig. \ref{fig:ebkmax} we see that the value \mbox{$K_{\rm max}=30$} provides a well converged ground state with energy 
\mbox{$\varepsilon_B =$ --1.5736 MeV}  in agreement with Ref.~\cite{Tilley04}. Assuming that the $\alpha$ particle matter and charge radii 
are 1.47 and 1.6755 fm, respectively, for the $^9$Be ground state we obtain a charge radius of \mbox{$r_{\text{ch}} =$ 2.508 fm} and a matter radius of \mbox{$r_{\text{mat}} =$ 2.466 fm}. 

Our value for the charge radius is in agreement with the experimental value of $2.519\pm0.012$ fm~\cite{Angeli13}. This 
reveals that our description of the system is rather accurate. For the matter radius our value is larger than the one given in Ref.~\cite{Tanihata88}, $2.38\pm0.01$ fm, 
obtained with Glauber-model calculations from interaction cross sections at high energies. A different estimation from a simple microscopic model by using cross sections at intermediate energies gives a radius of $2.53\pm0.07$ fm~\cite{Liatard90}, in better 
agreement with our calculation. It has been pointed out~\cite{Khalili96} that the optical limit approximation of Glauber models, such as in Ref.~\cite{Tanihata88}, may underestimate the radius of loosely bound systems. In halo nuclei, the few-body structure 
implies strong spatial correlations between the core and valence nucleons, so the optical limit fails. $^9$Be is not a halo system but it shows a strong few-body intrinsic configuration with the two $\alpha$ particles loosely bound by the remaining neutron, so 
the usual estimations of its radius from interaction cross sections may be misleading. 

\begin{table}
 \begin{tabular}{ccccc} 
 \toprule
 $i_{\rm max}$ & $\varepsilon_B$ (MeV) & $r_{\text{mat}}$ (fm) & $r_{\text{ch}}$ (fm) &$S_T(E1)$ (e$^2$fm$^2$) \\
 \colrule
 5  & $-$1.5659 & 2.453 & 2.502 &0.5565\\
 10 & $-$1.5734 & 2.465 & 2.507 &0.5760\\
 15 & $-$1.5736 & 2.466 & 2.508 &0.5762\\
 20 & $-$1.5736 & 2.466 & 2.508 &0.5762\\
 25 & $-$1.5736 & 2.466 & 2.508 &0.5762\\
% 30 & $-$1.5736 & 2.466 & 0.5762\\
 \botrule
 \end{tabular}
\caption{Ground-state energy $\varepsilon_B$, matter radius $r_{\text{mat}}$, charge radius $r_{\text{ch}}$ and sum rule $S_T(E1)$ as a function of $i_{\rm max}$ with $K_{\rm max} = 30$. A fast convergence is observed.}
\end{table}

In Table II we show the convergence of the ground-state energy, its matter radius, the charge radius and the sum rule for electric dipolar transitions (Eq. (\ref{eq:sumrule})) as the number of hyperradial 
excitations $i_{\rm max}$ increases. Calculations are performed for a fixed value of $K_{\rm max} = 30$, and we can see a rapid convergence.

%Another observable that is well determined experimentally for $^9$Be is the quadrupole moment of 5.29$\pm$0.04 e fm$^2$~\cite{sundholm91}
In addition, the $^9$Be system shows a large experimental quadrupole deformation, with a quadrupole moment of 5.29$\pm$0.04 e fm$^2$~\cite{sundholm91}. Our model provides a good description of this deformation due to the alpha-alpha cluster configuration, and 
gives a quadrupole moment of 4.91 e fm$^2$, which is close to the experimental value. 

\vspace{-10pt}
 \subsection{1/2$^+$ resonance}
 
The structure of the 1/2$^+$ resonance in $^9$Be has been studied by many authors, both theoretically~\cite{Efros99,Gorres95,Garrido10} and 
experimentally~\cite{Angulo99,Sumiyoshi02,Utsunomiya01,Arnold12,Burda10}, due to its relevance for the synthesis of this nucleus in Astrophysics. It has been found that the radiative capture 
reaction $\alpha(\alpha n,\gamma)^9\text{Be}$ is mainly governed by the 1/2$^+$ contribution of electric dipolar transitions to the ground state~\cite{Arnold12,Garrido11}. 

To get a well-defined $B(E1)$ distribution at low energies,
we need a basis with a large hyperradial
extension to concentrate many eigenvalues close to the breakup
threshold. For this purpose we describe the 1/2$^+$ states with a THO basis defined by parameters $b=0.7$ fm and $\gamma=0.7$ fm$^{1/2}$. %This choice enhances significantly the number of states near the threshold for a given $K_{\rm max}$ and a given number of hyperradial excitations $i_{\rm max}$.

\begin{figure}
\includegraphics[width=\linewidth]{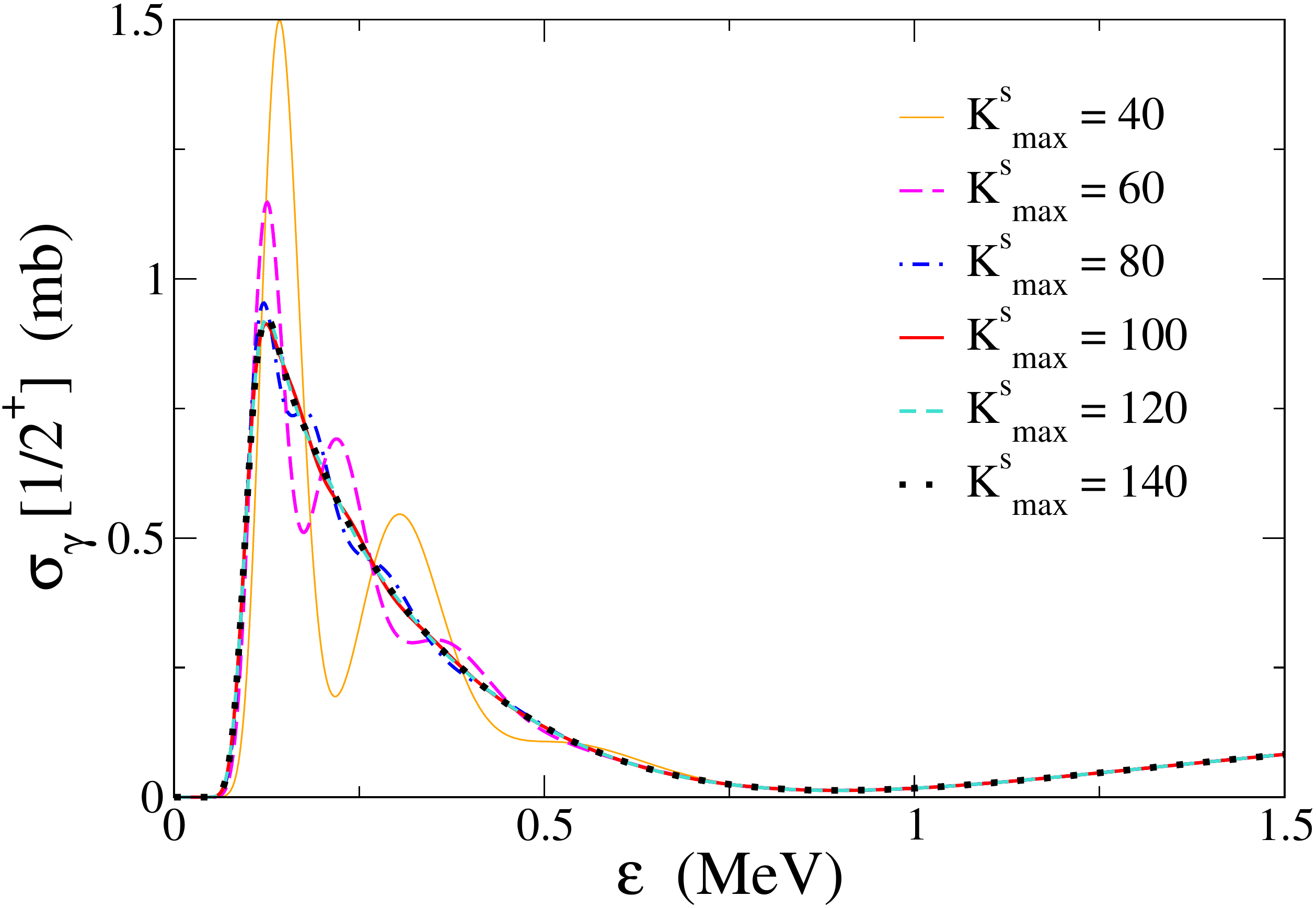}
\caption{(Color online) Dependence of the 1/2$^+$ contribution to the $^{9}$Be photodissociation cross section on $K_{\rm max}^s$. (See the text)}
\label{fig:resxsec}
\end{figure}

However, our calculations show a very slow convergence with respect to $K_{\rm max}$ for the low-energy 1/2$^+$ continuum. The structure of the 1/2$^+$ resonance is not well described with $K_{\rm max}$ values around 30-40, and going to larger hypermomenta involves 
the computation of very large basis sets, which is limited by computer power and calculation times. Since the 1/2$^+$ resonance decay is known to proceed mainly through the two-body low-lying s-wave $^8$Be resonance~\cite{Garrido10}, we expect the three-body 
resonance to be mainly governed by $\alpha$-$\alpha$ s-wave components. 
Thus we fix $K_{\rm max}$ to 40 and increase the maximum hypermomentum for s-waves, $K_{\rm max}^s$.
In Fig. \ref{fig:resxsec} we show the 1/2$^+$ contribution to the total photodissociation cross section, as a function of $K_{\rm max}^s$. 
For these calculations, we take a THO basis with $i_{\rm max}=30$ and we smooth the discrete values using Poisson distributions with a width parameter $w=30$. 
We can see in Fig. \ref{fig:resxsec} that the structure of the resonance is strongly dependent on $K_{\rm max}^s$ and very large values are needed to reach convergence. For this reason, we fix $K_{\rm max}^s$ to 140, maintaining the global $K_{\rm max}=40$ for all the other 
partial waves, as we find no need to include more channels in the wave function expansion to achieve converged cross section and reaction rates. The three-body force parameters needed to reproduce the position of the resonance are $v_{3b}=-2.45$ MeV, $r_{3b}=6.1$ fm, and $a_{3b}=5$.

\vspace{-10pt}
 \subsection{3/2$^+$, 5/2$^+$, 1/2$^-$ and 5/2$^-$ states}

 The 3/2$^+$, 5/2$^+$, 1/2$^-$ and 5/2$^-$ resonances in $^9$Be have excitation energies of 3.131, 1.475, 1.206 and 0.856 MeV, respectively~\cite{Tilley04}. Since these resonances contribute to the photodissociation cross section at higher energies than 
the case of 1/2$^+$, we expect smaller influences on the total reaction rate, at least in the low-temperature tail. We describe these states with a THO basis defined 
by parameters $b=0.7$ fm and $\gamma=1.0$ fm$^{1/2}$, that ensures enough states  at low energies. We include all channels up to $K_{\rm max}=30$, large enough to get converged strength distributions in these cases, and $i_{\rm max}=30$.
In order to adjust the position of the resonances, we change the parameter $v_{3b}$ to --1.60 MeV for the 3/2$^+$ states, --0.18 MeV for the 5/2$^+$, +1.65 MeV for the 5/2$^-$ states and +0.20 for the 1/2$^-$ states.
The $B(E1)$ and $B(M1)$ discrete values are smoothed using Poisson distributions with a width parameter $w=30,60,30$ for $3/2^+,5/2^+,1/2^-$, respectively. For the $5/2^-$ states we need a larger width parameter, which produces narrower distributions, since the $5/2^-$ 
resonance shows a very small width. This was previously reported in Ref.~\cite{MRoGa05}, where a value of $w=1300$ was used to describe properly the width of the narrow $2^+$ resonance in $^6$He. Thus we fix $w=10000$ around the resonance energy for the $5/2^-$ states, keeping 
$w=30$ for the non-resonant region.

Note that the convergence problem shown in the preceding subsection for the 1/2$^+$ state is absent in these cases. These resonances have larger excitation energies, and thus their properties are less sensitive to the 
 $\alpha$-$\alpha$ s-wave contribution.

\begin{figure}[!t]
\includegraphics[width=\linewidth]{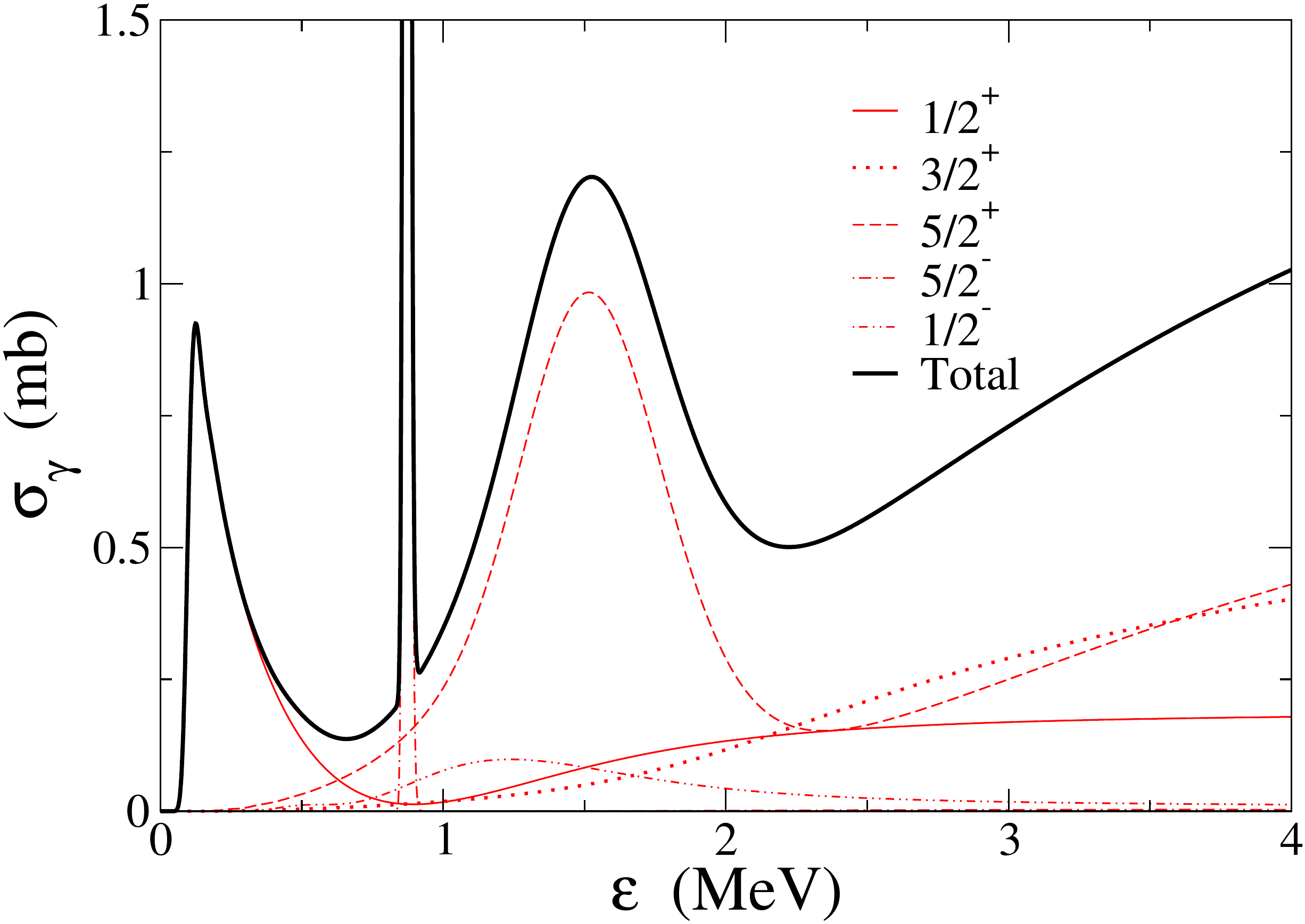}
\caption{(Color online) Contribution of the 1/2$^+$ (thin solid), 5/2$^+$ (dashed), 3/2$^+$ (dotted), 5/2$^-$ (dot dashed) and 1/2$^-$ (double dot dashed) states to the total photodissociation cross section (thick solid).}
\label{fig:xsec_3}
\end{figure}  

 \subsection{Photodissociation cross section}\label{ss:photo}

In Fig. \ref{fig:xsec_3}, we show the three electric dipolar contributions to the photodissociation cross section of $^9$Be from  1/2$^+$ (thin full line), 3/2$^+$ (dotted line), and 5/2$^+$ (dashed line) states. We include also the magnetic dipolar contribution 
from the 5/2$^-$ states (dot dashed) and the 1/2$^-$ states (double dot dashed). The total cross section is also shown by a thick full line. We can see how at very low energy only the 1/2$^+$ states contribute to the cross section. 

\begin{figure}
\includegraphics[width=\linewidth]{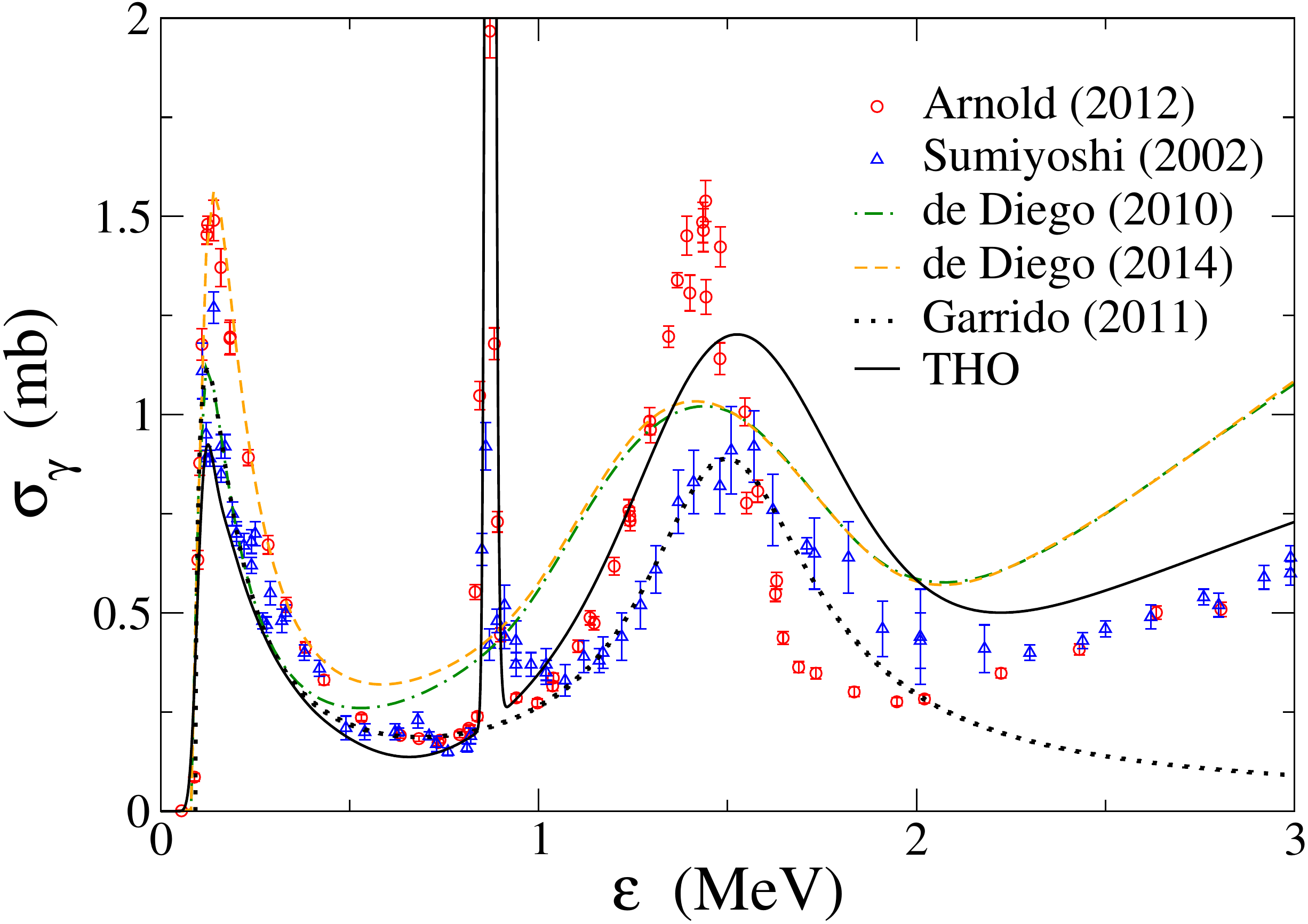}
\caption{(Color online) Total photodissociation from our three-body calculation (solid line) compared with the results from Ref.~\cite{RdDiego14} (dashed line),~\cite{RdDiego10} (dot dashed),~\cite{Garrido11} (dotted) and experimental data of 
Refs.~\cite{Sumiyoshi02} (triangles) and~\cite{Arnold12} (circles).}
\label{fig:xsec}
\end{figure}

In Fig. \ref{fig:xsec}, we compare the result shown in  Fig.~\ref{fig:xsec_3} with the experimental data from Arnold \textit{et al.}~\cite{Arnold12} and Sumiyoshi \textit{et al.}~\cite{Sumiyoshi02}. The agreement is rather good. Although we do not 
include them in the figure for clarity, our result is also in good agreement with other experimental data available in the literature~\cite{Burda10,Utsunomiya01}. We show also recent 
calculations by de Diego \textit{et al.}~\cite{RdDiego10,RdDiego14} using a similar three-body model. 
%However, the authors  were not able to describe the 1/2$^+$ resonant shape by using the finite energy interval approximation to generate the transition probability distribution. 
In these works, the continuum problem is solved by imposing box boundary conditions, for which obtaining a large density of states at the lowest energies is numerically challenging.
So, the 1/2$^+$ resonance peak for energies below 1.2 MeV is replaced by an energy-dependent Breit-Wigner distribution with the proper resonance parameters to reproduce the data. In Ref.~\cite{RdDiego10}, 
the 1/2$^+$ parameters are adjusted to reproduce the 2002 data, while in 
Ref.~\cite{RdDiego14} are fixed to describe the 2012 data. This procedure is applied by E. Garrido \textit{et al.}~\cite{Garrido11} to fit the total cross section 
including Breit-Wigner distributions for the lowest $^9$Be resonances, and we also include this 
result in Fig \ref{fig:xsec}. This calculation is adjusted to reproduce the data from Sumiyoshi \textit{et al.}

In contrast, our calculated 1/2$^+$ peak is directly obtained by smoothing the transition strength following Eq.~(\ref{eq:poisson}), using a THO basis that concentrates a large density of states near the breakup threshold. In this sense, our model 
provides the first full three-body calculation of the $^9$Be photodissociation cross section in the whole energy range. We underestimate the experimental data for the 1/2$^+$ contribution 
(in particular compared to 2012 data), but it shows the right low-energy behavior and the corresponding tail of the 
resonance. The smaller height is not crucial when computing the reaction rate, an observable that ranges over many orders of magnitude as a function of the temperature, 
especially at the lowest temperatures where the rate is strongly governed by the cross section 
behaviour up to 0.1-0.2 MeV only.

We reproduce very well the narrow 5/2$^-$ resonance, although we know from sequential models that this 
contribution has a small influence on the total reaction rate~\cite{Burda10,Arnold12}. This contribution is not computed in Refs.~\cite{RdDiego10,RdDiego14,Garrido11}. 
Concerning the 5/2$^{+}$ broad resonance, our three-body estimations agree better with Sumiyoshi \textit{et al.}~\cite{Sumiyoshi02} than with those from the more recent experiment of Arnold \textit{et al.}~\cite{Arnold12}, in which a rather narrow peak is 
obtained. For that reason we fix the position of the 5/2$^+$ resonance to Sumiyoshi \textit{et al.} data. In the calculations by de Diego \textit{et al.},  
the 5/2$^+$ resonance is adjusted to the energy given by Sumiyoshi \textit{et al.}, however due to the smoothing procedure the maximum is shifted  to lower energy.

The 3/2$^+$ resonance plays a minor role and its contribution affects only in 
the high energy region. At these energies, our calculations agree better with both sets of experimental data than those by de Diego \textit{et al}. The overall difference between both calculations could be associated to the different discretization 
methods and different two-body potentials. We have also estimated the M1 contribution to the 1/2$^-$ states, which has a small effect on the cross section, as shown in Fig.~\ref{fig:xsec_3}.

As we can see in Fig.~\ref{fig:xsec}, although the overall behavior is very similar in both sets of experimental data, there are important discrepancies between them. 
The accuracy of these experiments could then be questioned, since 
experimental normalization factors may lead to very different results. In Refs.~\cite{Sumiyoshi02} and~\cite{Arnold12}, for instance, the energy and width of the 1/2$^+$ resonance are found to be the same, but with different gamma widths by a factor of 1.3. This 
results in a different height for the resonant peak. For that reason it is not trivial to find an explanation to the differences between theory and experiment. We must also consider that three-body models are an approximation to the actual many-body 
problem, and consequently there might be effects on the cross section that we are not considering explicitly, e.g. dynamical effects within the clusters, full antisymetrization problems, etc.  Both calculations (this work and Refs.~\cite{RdDiego10,RdDiego14}) are 
systematically 
above the data at energies larger than 2 MeV, but at this level it is not possible to determine if this difference is related to many-body corrections or a possible normalization uncertainty. In any case, the final reaction 
rate at low temperature depends mainly on the photodissociation cross section at the lowest energies (0-0.2MeV) and the total strength, so diferences in the height, shape, etc of the specific structures are not crucial.

 \subsection{Reaction rate}

\begin{figure}
\includegraphics[width=\linewidth]{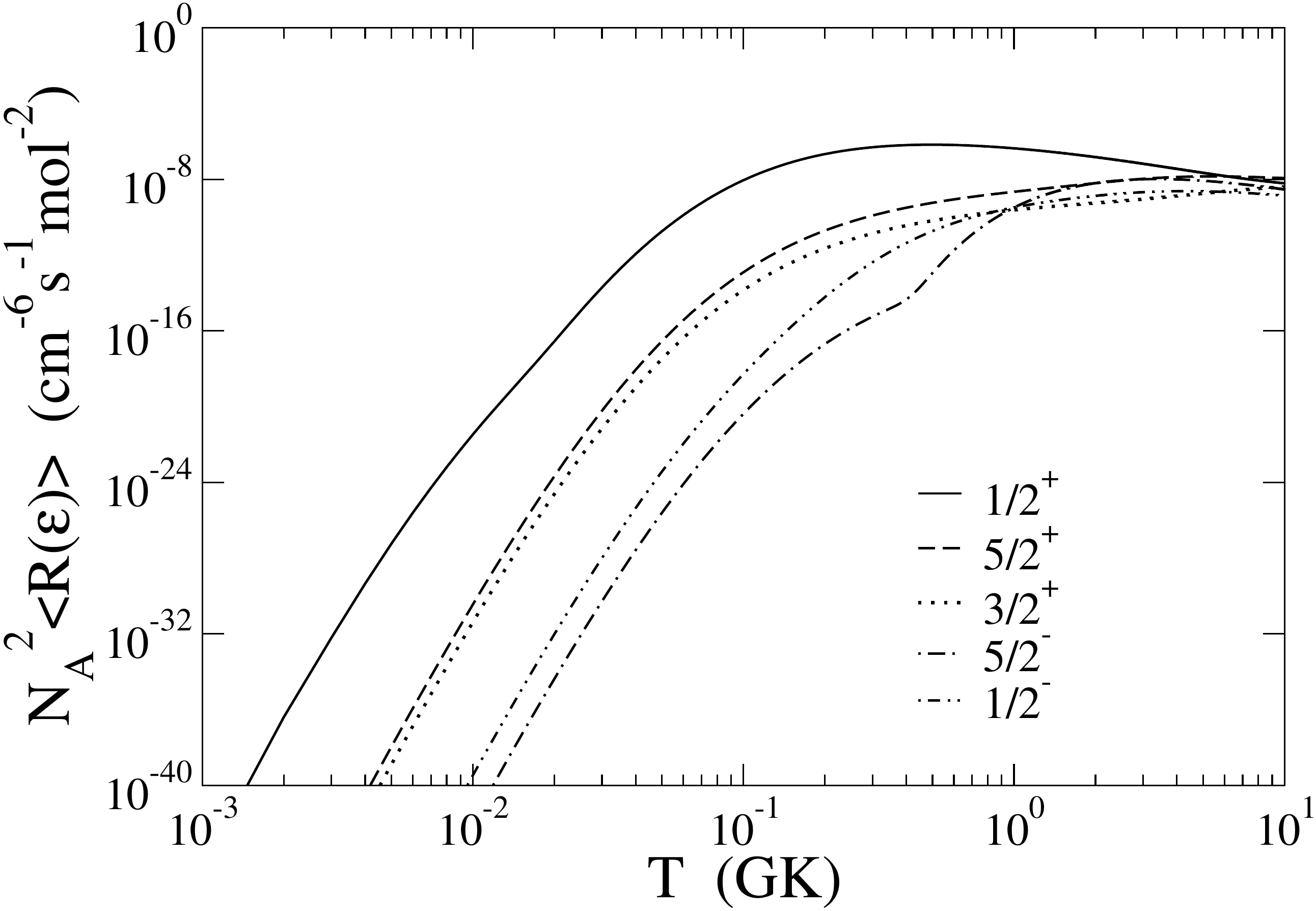}
\caption{Contribution of the 1/2$^+$ (thin solid), 5/2$^+$ (dashed), 3/2$^+$ (dotted), 5/2$^-$ (dot dashed) and 1/2$^-$ (double dot dashed) states to the total reaction rate.}
\label{fig:rates_3}
\end{figure}

We compute the rate of the radiative capture reaction $\alpha+\alpha+n\rightarrow$ $^9\text{Be} +\gamma $ from the photodissociation cross section, according to Eq.~(\ref{eq:aRE}). In Fig.~\ref{fig:rates_3} we show the 
contributions from the 1/2$^+$ (solid line), 3/2$^+$ (dotted line), 5/2$^+$ (dashed line), 5/2$^-$ (dot dashed) and 1/2$^-$ (double dot dashed) states to the reaction rate. We can see that the 1/2$^+$ states dominates over all other contributions, especially in 
the low-temperature tail of the reaction rate. The other contributions become relevant at temperatures above 3 GK.

\begin{figure}
\includegraphics[width=\linewidth]{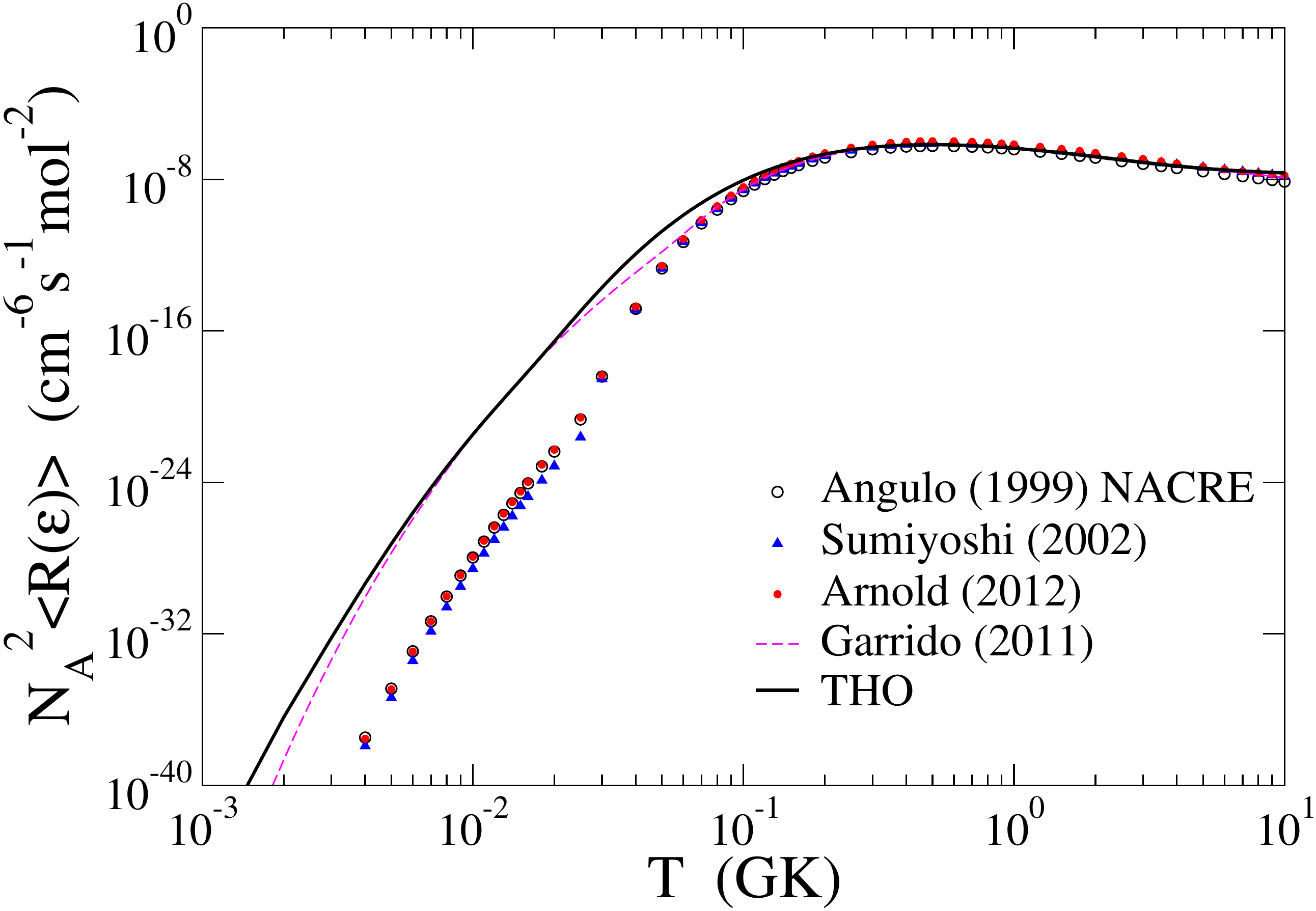}
\caption{(Color online) Total reaction rate from our three-body calculation (solid line) and three-body Breit-Wigner~\cite{Garrido11} (dashed line) compared with sequential estimations from experimental data of Refs.~\cite{Angulo99} (squares), \cite{Sumiyoshi02} (triangles), 
and \cite{Arnold12} (circles).} %\cite{Burda10} (squares)
\label{fig:rates}
\end{figure}

In Table~III we present the total reaction rate, the sum of the electric and magnetic dipolar contributions, at representative temperatures. In Fig.~\ref{fig:rates} we compare this rate with sequential estimations from experimental cross 
sections~\cite{Angulo99,Sumiyoshi02,Arnold12}. Our three-body model converges to the sequential result at high temperature, where the direct capture plays a minor role. Calculations by de Diego \textit{et al.}~\cite{RdDiego10,RdDiego14} between 0.1 GK and 5 GK 
also agree with this results, although we do not include them in Fig.~\ref{fig:rates} for clarity. At low temperature, below 0.1 GK, the three-body capture enhances the reaction rate in 
several orders of magnitude, in good agreement with three-body Breit-Wigner estimations by Garrido \textit{et al.}~\cite{Garrido11}. This confirms that the uncertainty related to the 1/2$^+$ resonance peak is not crucial when computing the reaction 
rate, as discussed in Subsection~\ref{ss:photo}. At such low temperatures the three-body system has no energy to populate the two-body $^8$Be resonance and, as expected, the 
direct capture begins to dominate. This effect cannot be described with sequential models. %Burda10

\setlength{\tabcolsep}{4pt}
\begin{table}\label{tabla}
 \begin{tabular}{llllll} 
 \toprule
 $T_9$  & Rate & $T_9$  & Rate & $T_9$  & Rate \\
 \colrule
 0.001 & 3.67$\times10^{-45}$ & 0.04  & 1.16$\times10^{-12}$ & 0.45  & 6.78$\times10^{-7}$ \\ 
 0.002 & 4.03$\times10^{-37}$ & 0.05  & 1.83$\times10^{-11}$ & 0.5   & 6.85$\times10^{-7}$ \\ 
 0.003 & 6.19$\times10^{-33}$ & 0.06  & 1.31$\times10^{-10}$ & 0.6   & 6.61$\times10^{-7}$ \\ 
 0.004 & 4.57$\times10^{-30}$ & 0.07  & 5.71$\times10^{-10}$ & 0.7   & 6.10$\times10^{-7}$ \\ 
 0.005 & 5.75$\times10^{-28}$ & 0.08  & 1.78$\times10^{-9}$ & 0.8   & 5.52$\times10^{-7}$ \\ 
 0.006 & 2.48$\times10^{-26}$ & 0.09  & 4.38$\times10^{-9}$  & 0.9   & 4.94$\times10^{-7}$ \\ 
 0.007 & 5.17$\times10^{-25}$ & 0.1   & 9.07$\times10^{-9}$  & 1     & 4.41$\times10^{-7}$ \\ 
 0.008 & 6.41$\times10^{-24}$ & 0.11  & 1.65$\times10^{-8}$  & 1.25  & 3.32$\times10^{-7}$ \\ 
 0.009 & 5.41$\times10^{-23}$ & 0.12  & 2.71$\times10^{-8}$  & 1.5   & 2.53$\times10^{-7}$ \\ 
 0.011 & 3.40$\times10^{-22}$ & 0.13  & 4.11$\times10^{-8}$  & 1.75  & 1.98$\times10^{-7}$ \\ 
 0.012 & 7.15$\times10^{-21}$ & 0.14  & 5.85$\times10^{-8}$  & 2     & 1.58$\times10^{-7}$ \\ 
 0.013 & 2.61$\times10^{-20}$ & 0.15  & 7.93$\times10^{-8}$  & 2.5   & 1.07$\times10^{-7}$ \\ 
 0.014 & 8.55$\times10^{-20}$ & 0.16  & 1.03$\times10^{-7}$  & 3     & 7.89$\times10^{-8}$ \\ 
 0.015 & 2.57$\times10^{-19}$ & 0.17  & 1.29$\times10^{-7}$  & 3.5   & 6.18$\times10^{-8}$ \\ 
 0.016 & 7.25$\times10^{-19}$ & 0.18  & 1.57$\times10^{-7}$  & 4     & 5.09$\times10^{-8}$ \\ 
 0.017 & 1.93$\times10^{-18}$ & 0.19  & 1.87$\times10^{-7}$  & 5     & 3.85$\times10^{-8}$ \\ 
 0.018 & 4.91$\times10^{-18}$ & 0.2   & 2.18$\times10^{-7}$  & 6     & 3.19$\times10^{-8}$ \\ 
 0.019 & 1.19$\times10^{-17}$ & 0.25  & 3.72$\times10^{-7}$  & 7     & 2.79$\times10^{-8}$ \\ 
 0.02  & 2.79$\times10^{-17}$ & 0.3   & 5.01$\times10^{-7}$  & 8     & 2.52$\times10^{-8}$ \\ 
 0.025 & 1.11$\times10^{-15}$ & 0.35  & 5.93$\times10^{-7}$  & 9     & 2.34$\times10^{-8}$ \\ 
 0.03  & 1.95$\times10^{-14}$ & 0.4   & 6.49$\times10^{-7}$  & 10    & 2.20$\times10^{-8}$ \\
 \botrule
 \end{tabular}
\caption{Reaction rate of $\alpha\alpha n$, in cm$^{-6}$s$^{-1}$mol$^{-2}$, at representative temperatures in GK, $T_9$.}
\end{table}

 \section{Summary and conclusions}\label{sec:conclusions}

The structure of the Borromean nucleus $^9$Be ($\alpha+\alpha+n$) has been described in a full three-body model using the analytical THO method. The photodissociation cross section is calculated including electric dipolar transitions from the $3/2^-$ 
ground state to the $1/2^+$, $3/2^+$, $5/2^+$ continuum states and also magnetic transitions to the $5/2^-$ and $1/2^-$ states. For each angular momentum, an appropriate analytical THO basis has been used. The results show the dominance of the 1/2$^+$ resonance 
at low energy. The comparison with the experimental data and with previous calculations available in the literature reveals the goodness of the formalism. 

The difference between theoretical works is discussed. Unlike previous calculations, our model describes the photodissociation cross section using the same footing in the whole energy range. The differences between theory and experiments might be 
related to many-body corrections not included within three-body models and also to experimental uncertainties arising from the discrepancies between the different data sets.
%Previous calculations available in the literature do not describe the 1/2$^+$ contribution to the photodissociation cross section directly in a three-body model. Calculations by de Diego \textit{et al.} and Garrido \textit{et al.} fit the  experimental 
%1/2$^+$ resonant shape with an energy dependent Breit-Wigner. Our three-body model describes the photodissociation cross section in the whole energy range. The differences between theory and experiments is discussed.

The radiative capture reaction rate  for the formation of $^9$Be is then calculated from the photodissociation cross section. The reaction rate so obtained within a full three-body model matches the reaction rates obtained using sequential models at temperatures 
above 0.1 GK. However at lower temperatures the three-body calculation is several orders of magnitude larger than the sequential models. This result reveals the sequential models fail to reproduce the capture reaction rate of $^9$Be at low temperature where the 
three-body system has no energy to populate the two-body $^8$Be resonance and the direct capture becomes more relevant. Our calculations agree reasonably well with estimations using three-body Breit-Wigner distributions to fit the cross section.

The successful application of the analytical THO method to the determination of the $^9$Be photodissociation cross section and radiative capture reaction rate encourages the application to the triple-alpha process as well as the formation of $^{17}$Ne ($^{15}$O+$p$+$p$). Both reactions
involve three charged particles, which increase the level of difficulty.

\begin{acknowledgments}
Authors are grateful to  P. Descouvemont, E. Garrido and J. Gómez-Camacho for useful discussions and suggestions. 
This work has been partially supported by the Spanish Ministerio de Econom\'{\i}a y Competitividad under Projects FPA2009-07653 and FIS2011-28738-c02-01, by Junta de Andaluc\'{\i}a under group number FQM-160 and Project P11-FQM-7632, and by the
Consolider-Ingenio 2010 Programme CPAN (CSD2007-00042). 
This work was performed under the auspices of the U.S. Department of Energy by Lawrence Livermore National Laboratory under Contract DE-AC52-07NA27344.
J. Casal acknowledges a FPU research grant from the Ministerio de Educaci\'on, Cultura y Deporte.
\end{acknowledgments}

\vspace{5pt}
\appendix
\section{Magnetic operator matrix elements}\label{appendix:1}

In this Appendix, we present the main expressions needed to compute the magnetic operator matrix elements from Eq.~(\ref{eq:Mop}). For each particle $q$, we rotate the wave function given by Eq.~(\ref{eq:wf}) to the Jacobi-$q$ system, and then we sum up the 
orbital and spin contributions. This can be expressed in a compact form by using the transformations between different Jacobi sets for the angular part of the wave functions~\cite{IJThompson04},
\begin{equation}
 N_{\beta_k\beta_q}=\langle k:\beta_kj\mu|q:\beta_qj\mu\rangle.
 \label{eq:trans}
\end{equation}
Here index $q$ labels particle $q$, while $k$ denotes the preferred Jacobi system in which we diagonalize the Hamiltonian. Since $k$ is fixed, we omit it for the following expressions, so $|k:\beta_kj\mu\rangle$ represents $\mathcal{Y}_{\beta j\mu}(\Omega)$ in 
Eq.~(\ref{eq:basis}). These functions are expanded in hyperspherical harmonics (HH)~\cite{Zhukov93,MRoGaTh} $\Upsilon_{Klm_l}^{l_xl_y}(\Omega)$ as 
\begin{eqnarray}
\mathcal{Y}_{\beta j\mu}(\Omega)& =& \sum_{\nu\iota}\langle j_{ab}\nu I\iota|j\mu\rangle \kappa_I^{\iota} \nonumber \\
&\times&  \sum_{m_l\sigma}\langle lm_lS_x\sigma|j_{ab}\nu\rangle
\Upsilon_{Klm_l}^{l_xl_y}(\Omega)\chi_{S_x}^{\sigma}.
\label{eq:HHexpand}
\end{eqnarray}
Here $\chi_{S_x}^{\sigma}$ is the spin wave function of the two particles related by the Jacobi
coordinate $\boldsymbol{x}$, and $\kappa_I^{\iota}$ is the spin
function of the third particle.The HH are eigenfunctions of the hypermomentum operator $\widehat{K}^2$, and can be expressed in terms of the spherical harmonics as 
\begin{eqnarray}
\Upsilon_{Klm_l}^{l_xl_y}(\Omega)&=&\sum_{m_{x}m_{y}}\langle l_xm_xl_ym_{y}|lm_l \rangle\nonumber\\ &\times&\Upsilon_{K}^{l_xl_ym_{x}m_{y}}(\Omega),\\
 \Upsilon_K^{l_x l_y m_x m_y}(\Omega)&=&\varphi_K^{l_x l_y}(\alpha)
 Y_{l_x m_x}(\widehat{x}) Y_{l_y m_y}(\widehat{y}),\\ 
 \label{eq:HH}
\varphi_K^{l_x l_y}(\alpha) &=& N_K^{l_x l_y} (\sin\alpha)^{l_x}
(\cos\alpha)^{l_y} \nonumber \\
&\times& P_n^{l_x+\frac{1}{2},l_y+\frac{1}{2}}(\cos 2\alpha),
\end{eqnarray}
where $P_n^{a,b}$ is a Jacobi polynomial with order $n=(K-l_x-l_y)/2$ and $N_K^{l_x l_y}$ is the normalization constant.

Using Eq.~(\ref{eq:trans}) and expanding the explicit angular dependence of the wave functions, we can express the orbital and spin parts of the magnetic operator reduced matrix element for a given multipolarity $\lambda$ as 
%Eq.~(\ref{eq:me_orb}) and~(\ref{eq:me_spin}), respectively.
\begin{widetext}
\begin{eqnarray}
 \nonumber\langle nj||\widehat{M}_\lambda^{\text{orb}}||n'j'\rangle & = &\frac{e\hbar}{2mc}\frac{\sqrt{\lambda}}{\lambda+1}\hat{(\lambda-1)}\hat{\lambda}\hat{j}'(-1)^{\lambda} \sum_q\left(\frac{M_T-m_q}{M_T}\right)^{\lambda}\left(\frac{m}{a_{y_q}}\right)^{\frac{\lambda-1}{2}} 2g_l^{(q)}\sum_{\beta\beta'}\sum_{\beta_q\beta_q'}N_{\beta\beta_q}N_{\beta\beta_q'}\delta_{S_{x_q}S_{x_q}'}\delta_{l_{x_q}l_{x_q}'}\\
 &&\nonumber \times (-1)^{2j-j'+l'_{y_q}-l_{y_q}+l_{x_q}-S_{x_q}+j_{ab_q}+j'_{ab_q}-I_q}\sqrt{l'_{y_q}\left(l'_{y_q}+1\right)}\hat{l}_{y_q}\hat{l}_{y_q}'^2\hat{j}_{ab_q}\hat{j}'_{ab_q}\hat{l}_{q}\hat{l}'_{q}\tj{l_{y_q}}{\lambda-1}{l'_{y_q}}{0}{0}{0}\\
 &&\nonumber \times W(l_{y_q}l'_{y_q}(\lambda-1)1;\lambda l'_{y_q})W(l_{q}l'_{q}l_{y_q}l'_{y_q};\lambda l_{x_q})W(l_{y_q}l'_{y_q}(\lambda-1)1;\lambda l'_{y_q})W(l_{y_q}l'_{y_q}(\lambda-1)1;\lambda l'_{y_q})\\
 && \times \sum_{ii'}C_n^{i\beta j}C_n'^{i'\beta'j'}\int\int d\alpha d\rho(\sin\alpha)^2(\cos\alpha)^2 U_{i\beta}(\rho)\varphi_{K_q}^{l_{x_q}l_{y_q}}(\alpha)y^{\lambda-1}U_{i'\beta'}(\rho)\varphi_{K'_q}^{l'_{x_q}l'_{y_q}}(\alpha).
\label{eq:me_orb}
\end{eqnarray}
\begin{eqnarray}
 \nonumber\langle nj||\widehat{M}_\lambda^{\text{spin}}||n'j'\rangle & = &\frac{e\hbar}{2mc}\sqrt{\lambda}\hat{(\lambda-1)}\hat{\lambda}\hat{j}' \sum_q\left(\frac{M_T-m_q}{M_T}\right)^{\lambda-1}\left(\frac{m}{a_{y_q}}\right)^{\frac{\lambda-1}{2}} g_s^{(q)}\sum_{\beta\beta'}\sum_{\beta_q\beta_q'}N_{\beta\beta_q}N_{\beta\beta_q'}\delta_{S_{x_q}S_{x_q}'}\delta_{l_{x_q}l_{x_q}'}\\
 &&\nonumber \times (-1)^{j+j'+l_{x_q}-S_{x_q}-j_{ab_q}+2I_q}\sqrt{I_{q}\left(I_{q}+1\right)}\hat{I}_q\hat{l}_{y_q}\hat{l}_{y_q}'\hat{j}_{ab_q}\hat{j}_{ab_q}'\hat{l}_{q}\hat{l}_{q}'\tj{l_{y_q}}{\lambda-1}{l'_{y_q}}{0}{0}{0}\\
 &&\nonumber \times W(l_{q}l'_{q}l_{y_q}l'_{y_q};(\lambda-1) l_{x_q})W(l_{q}l'_{q}j_{ab_q}j'_{ab_q};(\lambda-1) S_{x_q})\left\{\begin{array}{ccc}j&j'&\lambda\\j_{ab_q}&j'_{ab_q}&\lambda-1\\I_q&I_q&1\end{array}\right\}\\
 && \times \sum_{ii'}C_n^{i\beta j}C_n'^{i'\beta'j'}\int\int d\alpha d\rho(\sin\alpha)^2(\cos\alpha)^2 U_{i\beta}(\rho)\varphi_{K_q}^{l_{x_q}l_{y_q}}(\alpha)y^{\lambda-1}U_{i'\beta'}(\rho)\varphi_{K'_q}^{l'_{x_q}l'_{y_q}}(\alpha).
\label{eq:me_spin}
\end{eqnarray}
\end{widetext}

The notation $\hat{j}$ represents a reduced form for the factor $\sqrt{2j+1}$. These expressions depend on the orbital and spin $g$-factors of each particle. The $\alpha$ particles have spin zero, so we consider $g_s^{(\alpha)}=0$ and $g_l^{(\alpha)}$ is taken as its 
charge. For the neutron we use the free value of $g_s^{(n)}=-3.82$ and we do not assign any effective charge, so $g_l^{(n)}=0$. It is known that the effective $g$-factor are rather uncertain~\cite{Romero08}, especially $g_s^{(n)}$ which could be reduced by a 
factor of 2 due to spin polarization. A more exhaustive analysis of these factors for the particular case of $^9$Be could reduce the uncertainty in the magnetic contributions to the photodissociation cross section. 

\section{Smoothing procedure}\label{appendix:2}

\begin{figure}[!ht]
\includegraphics[width=\linewidth]{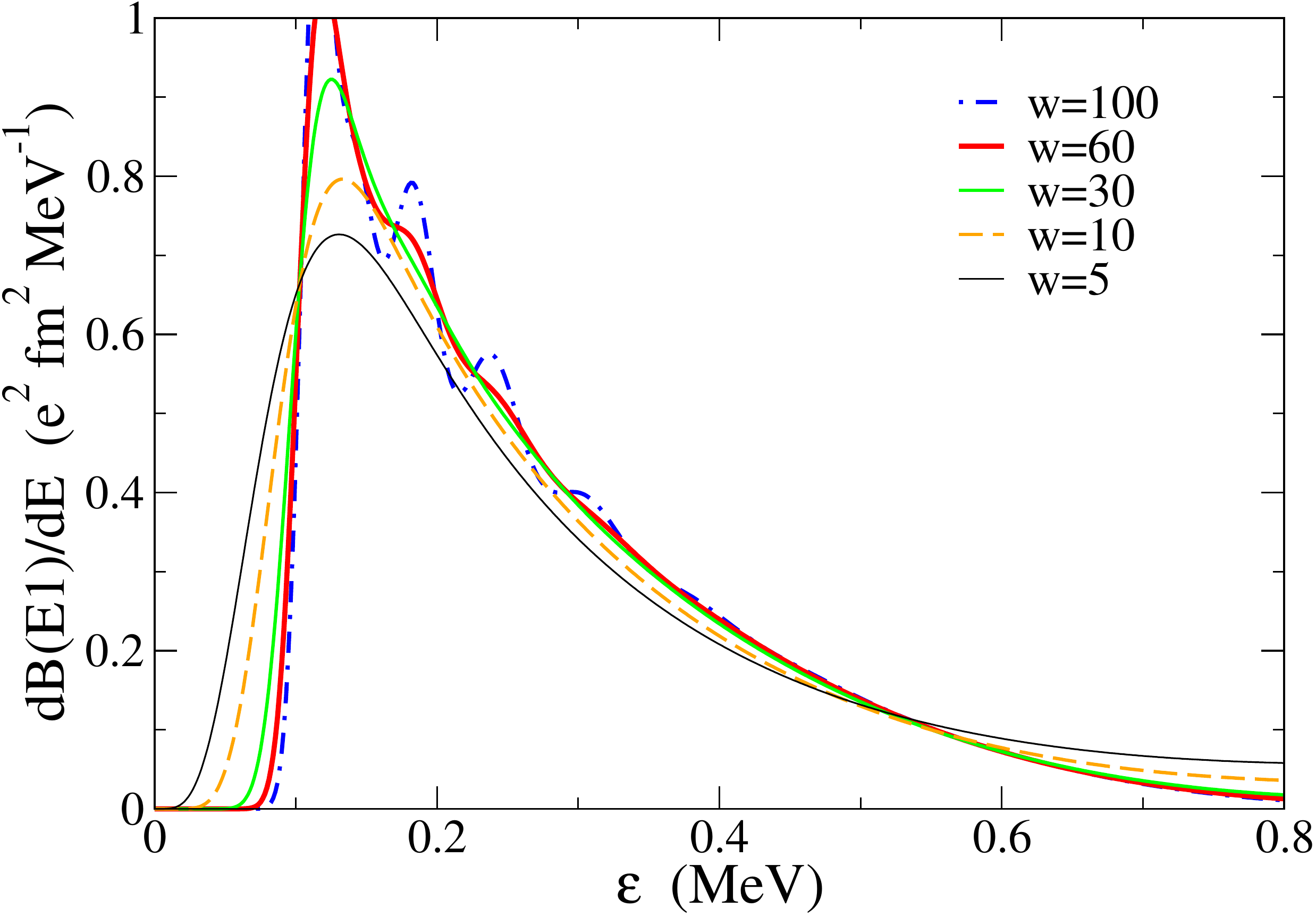}
\caption{(Color online)$B(E1)$ distribution to the 1/2$^+$ states as a function of the Poisson width parameter $w$. (See text for details).}
\label{fig:wappendix}
\end{figure}

In PS methods any transition probability to be calculated is given by a set of discrete values. In order to obtain a continuous distribution, in this work we assign a Poisson distributions to each PS. 
We discuss in this Appendix the procedure to select an optimal width parameter $w$ for the Poisson distributions defined by Eq.~(\ref{eq:poisson}). The value of $w$ must ensure a smooth $B(E1)$ distribution without spreading it unphysically.
As an example, we show in Fig.~\ref{fig:wappendix} the $B(E1)$ distribution to the 1/2$^+$ states calculated with different width parameters. For $w$ values smaller than 30, the distributions are too wide  to represent the PS energy distributions, and 
consequently the final distributions cannot reproduce the experimental photodissociation data. For much larger values, however, the final distributions are distorted and show unphysical oscillations or peaks. This is our prescription to select the optimal $w$ 
value, that is $w$ as large as possible. In this case $w=30$ is a reasonable choice. This method provides good results and a rather good agreement with the experimental data on the photodissociation cross section, as shown in Subsection~\ref{ss:photo}.

\newpage
\bibliography{./bibfile}

\end{document}